\documentstyle[prd,aps,floats,epsfig,graphicx]{revtex}

\def\bea{\begin{eqnarray}}
\def\eea{\end{eqnarray}}

\begin{document}
\draft 

%
%
 
%
\renewcommand{\topfraction}{0.99}
\renewcommand{\bottomfraction}{0.99}
\twocolumn[\hsize\textwidth\columnwidth\hsize\csname 
@twocolumnfalse\endcsname
  
\title
{\Large Brane Gases in the Early Universe}
\author{S. Alexander, R. Brandenberger and D. Easson} 
\address{~\\Department of Physics, Brown University,  
Providence, RI 02912, USA.}
\date{\today} 
\maketitle
\begin{abstract} 
Over the past decade it has become clear that fundamental strings are not 
the only fundamental degrees of freedom in string theory. D-branes are also 
part of the spectrum of fundamental states. In this paper we explore some 
possible effects of D-branes on early Universe string cosmology, starting with
two key assumptions: firstly that the initial state of the Universe 
corresponded to a dense, hot gas in which all degrees of freedom were in thermal equilibrium, and secondly that the topology of the background space admits one-cycles. We argue by t-duality that in this context the cosmological singularities are not present. We derive the equation of state of the brane gases and apply the results to suggest that, in an expanding background, the winding modes of fundamental strings will play the most important role at late times. In particular, we argue that the string winding modes will only allow four space-time dimensions to become large. The presence of brane winding modes with $p > 1$ may lead to a hierarchy in the sizes of the extra dimensions.        
\end{abstract}

\pacs{PACS numbers: 98.80Cq      \hspace{4.7cm} BROWN-HET-1221, hep-th/0005212}]

\vskip 0.4cm
\section{Introduction}

In this paper we consider an approach to string cosmology in close
analogy to the usual starting point of standard big-bang cosmology.
We assume that the Universe started out small, dense, hot, and with all
fundamental degrees of freedom in thermal equilibrium. We also
assume that the background space is toroidal in all spatial dimensions \footnote{In fact, we do not need to be this specific. The crucial assumption
is the existence of one-cycles in all spatial directions.}. Given these
assumptions, the initial state will consist of a gas of all fundamental branes
which the theory admits. We will study the equation of state of the individual types of branes, neglecting for simplicity brane interactions, and will use the results to determine the source terms in the equations of motion for the background. In
particular, we will study obstructions to spatial dimensions becoming large.
We will find that fundamental string winding modes dominate the evolution at
late times, and prevent more than three spatial dimensions from becoming large, in agreement with the scenario proposed in \cite{BV} in the context of perturbative string theory. We also argue that because of t-duality the usual  singularities of a the homogeneous and isotropic big bang and inflationary cosmologies are not present.

The main goal of this paper is to generalize the considerations of \cite{BV} to the context of our present understanding of string theory. In the 1980's, it was believed that the only fundamental degrees of freedom of string theory were the fundamental strings. In this context, it was shown that t-duality and string winding modes could have a very important effect on early Universe cosmology.
Assuming that the background space is toroidal and thus admits string winding modes, it was shown that t-duality could explain the absence of the initial big-bang singularity. In addition, it was speculated that string winding modes would only allow three
spatial dimensions to become large. The second point was put on a firmer 
basis by the work of Tseytlin and Vafa \cite{TV}, who discussed the 
effects of gases of strings on the background equations of motion which 
were taken to be those of dilaton gravity.

The first main point of the BV scenario \cite{BV} is that t-duality will lead to an equivalence of the physics if the radius of the background torus changes (in string units) from $R$ to $1/R$. This corresponds to an interchange of momentum and winding modes. Thus, $R$ becoming small is equivalent to $R$ tending to infinity. Neither limit corresponds to a singularity for string matter. For example, the temperature $T$ obeys
\begin{equation} \label{tempdual}
T({1 \over R}) \, = \, T(R) \, .
\end{equation}
Thus, in string cosmology the big bang singularity can be avoided. 
The second point suggested in \cite{BV} was that string winding modes 
would prevent more than three spatial dimensions from becoming large. 
The point is that string winding modes cannot annihilate in more than 
three spatial dimensions (by
a simple classical dimension counting argument). 
In the context of dilaton cosmology,
a gas of string winding modes (which has an equation of state 
$\tilde{p} = - (1/d) \rho$, where $\tilde{p}$ and $\rho$ denote 
pressure and energy density, respectively, and $d$ is the number of 
spatial dimensions) will lead to a confining
potential in the equation of motion for $\lambda = log(a)$, where $a(t)$ is the scale factor of the Universe \cite{TV}. Note that this is not the result which would be obtained in a pure metric background obeying the Einstein equations.
The dynamics of classical strings in higher dimensional expanding backgrounds was studied numerically in \cite{MS}, confirming the conclusions of \cite{BV}.

However, it is now clear that string theory has a much richer set of fundamental degrees of freedom, consisting - in addition to fundamental strings - of D-branes \cite{Pol} of various dimensionalities. The five previously known
consistent perturbative string theories are now known to be connected by a web of dualities \cite{Witten1}, and are believed to represent different corners of moduli space of a yet unknown theory called M-theory. Which branes arise in the effective string
theory description depends on the particular point in moduli space. We will be making a specific assumption below.

The question we would like to address is whether the inclusion of the new fundamental degrees of freedom will change the main cosmological implications of string theory suggested in \cite{BV}, namely the avoidance of the initial cosmological singularity, 
and the singling out of 3 as the maximal number of large spatial dimensions, in the context of an initial state which is assumed to be hot, dense and small, and of a background geometry which admits string winding modes.

Our concrete starting point is 11-dimensional M-theory compactified on $S^1$ to yield 10-dimensional Type II-A string theory. The resulting low energy effective theory is supersymmetrized dilaton gravity. As fundamental states, M-theory admits the graviton, 2-branes and 5-branes. After compactification, this leads to 0-branes, 1-branes, 2-branes, 4-branes, 5-branes, 6-branes and 8-branes as the fundamental extended objects of the 10-dimensional theory. The dilaton represents the radius of the 
compactified $S^1$. We are in a region of moduli space in which the string coupling constant $g_s$ is smaller than 1. 

We assume that all spatial dimensions are toroidal (radius $R$), and that the Universe starts out small, dense, hot, and in thermal equilibrium. Thus, the Universe will contain a gas of all branes appearing in the spectrum of the theory. Note that this starting point is in close analogy with the hot big bang picture in standard cosmology, but very different from brane-world scenarios in which the existence of a particular set of branes is postulated from the outset without much justification from the point of view of cosmology. 

There have been several interesting previous studies of the cosmology of brane gases. Maggiore and Riotto \cite{MR99} (see also \cite{R99}) studied the phase diagram of brane gases motivated by M-theory as a function of the string coupling constant and of
the Hubble expansion rate (as a measure of space-time curvature) and discovered regions of the phase diagram in which brane gases determine the dynamics, and regions in which the effective action is no longer well described by a ten-dimensional supergravity action. Given our assumptions, we are in a region in moduli space in which the ten-dimensional effective description of the physics remains true to curvature scales larger than that given by the string scale. In this paper, we consider the time evolution of the system through phase space starting from some
well-defined initial conditions. We will argue that as a consequence of t-duality, curvature scales where the ten-dimensional description breaks down are never reached. 

In another interesting paper, Park et al. \cite{PSL00} take a starting point very close to our own, a hot dense gas of branes. However, they did not consider the winding and oscillatory modes of the branes.

In the following section we will study the equation of state of the brane gases for all values of their spatial dimension p. We will separately analyze the contributions of winding and non-winding modes (the latter treated perturbatively). The results will be used as source terms for the equations of motion of the background dilaton gravity fields, following the approach of \cite{TV}. We find that the winding modes of any p-brane lead to a confining force which prevents the expansion of the spatial dimensions, and that the branes with the largest value of p give the largest contribution to the energy of the gas in the phase in which the scale factor is increasing.

In Section 3 we use the results of the previous section to argue that the main conclusions of the scenario proposed in \cite{BV} are unchanged: t-duality eliminates the cosmological singularity, and winding modes only allow three dimensions of space to become large. We point out a potential problem (the {\it brane problem}) of  cosmologies based on theories which admit branes in their spectrum of fundamental states. This problem is similar to the well-known domain wall problem \cite{problem} in cosmological models based on quantum field theory. It is pointed out that a phase of loitering (see e.g. \cite{SFS}) yields a natural solution of this problem, and it is shown that the background equations of motion may well yield a loitering stage during the early evolution of the Universe. Some limitations of our considerations and avenues for future research are discussed in the final section.

\vskip 0.4cm
\section{Equation of State of Brane Gases}

As mentioned in the Introduction, our starting point is Type II-A string 
theory on a 9-dimensional toroidal background space (with the time direction being infinite), resulting from the compactification of M-theory on $S^1$. The fundamental degrees of freedom are the fields of the bulk background (resulting from the graviton in M-theory), 1-branes, 2-branes, 4-branes, 5-branes, 6-branes and 8-branes.

The low-energy bulk effective action is given by
\begin{eqnarray} \label{bulk}
S_{bulk} \, = \, {1 \over {2 \kappa^2}} \int d^{10}x \sqrt{-G} e^{-2 \phi} \bigl[ R &+& 4 G^{\mu \nu} \nabla_\mu \phi \nabla_\nu \phi \nonumber \\
&-& {1 \over {12}} H_{\mu \nu \alpha}H^{\mu \nu \alpha} \bigr] \, ,
\end{eqnarray}
where $G$ is the determinant of the background metric $G_{\mu \nu}$, $\phi$ is the dilaton, $H$ denotes the field strength corresponding to the bulk antisymmetric tensor field $B_{\mu \nu}$, and $\kappa$ is determined by the 10-dimensional Newton constant in the usual way.

The total action is the sum of the above bulk action and the action of all branes present. The action of an individual brane with spatial dimension p has the Dirac-Born-Infeld form \cite{Pol}
\begin{equation} \label{brane}
S_p \, = \, T_p \int d^{p + 1} \zeta e^{- \phi} \sqrt{- det(g_{mn} + b_{mn} + 2 \pi \alpha' F_{mn})}
\end{equation}
where $T_p$ is the tension of the brane, $g_{mn}$ is the induced metric on the brane, $b_{mn}$ is the induced antisymmetric tensor field, and $F_{mn}$ the field strength tensor of gauge fields $A_m$ living on the brane. The total action is the sum of the 
bulk action (\ref{bulk}) and the sum of all of the brane actions (\ref{brane}), each coupled as a delta function source (a delta function in the directions transverse to the brane) to the 10-dimensional action. 

The induced metric on the brane $g_{mn}$, with indices $m,n,...$ denoting space-time dimensions parallel to the brane, is determined by the background metric $G_{\mu \nu}$ and by scalar fields $\phi_i$ (not to be confused with the dilaton $\phi$) living on the brane (with indices $i,j,...$ denoting dimensions transverse to the brane) which describe the fluctuations of the brane in the transverse directions:
\begin{equation} \label{indmet}
g_{mn} \, = \, G_{mn} + G_{ij} \partial_m \phi_i \partial_n \phi_j + G_{in} \partial_m \phi_i \, .
\end{equation}
The induced antisymmetric tensor field is
\begin{equation} \label{indten}
b_{mn} \, = \, B_{mn} + B_{ij} \partial_m \phi_i \partial_n \phi_j + B_{i[n} \partial_{m]} \phi_i \, .
\end{equation}
In addition,
\begin{equation}
F_{mn} \, = \, \partial_{[m} A_{n]} \, .
\end{equation}

In the string frame, the fundamental string has tension
\begin{equation}
T_f \, = \, 2 \pi \alpha' \, ,
\end{equation}
whereas the brane tensions for various values of p are given by \cite{Pol}
\begin{equation}
T_p \, = \, {{\pi} \over {g_s}} (4 \pi^2 \alpha')^{-(p + 1)/2} \, ,
\end{equation}
where $\alpha' \sim l_{st}^2$ is given by the string length scale $l_{st}$ and $g_s$ is the string coupling constant \footnote{Note that the s-duality between the fundamental and D-string tensions is more evident in the Einstein frame.}.
Note that all of these branes have positive tension.

In the following, we wish to compute the equation of state of the brane gases for a general value of p. For our considerations, the most important modes are the winding modes. If the background space is $T^9$, a p-brane can wrap around any set of p toroidal directions. The modes corresponding to these winding modes by t-duality are the momentum modes corresponding to center of mass motion of the brane. The next most important modes for our considerations are the modes corresponding to fluctuations of the 
brane in transverse directions. These modes are in the low-energy limit described by the brane scalar fields $\phi_i$. In addition, there are bulk matter fields and brane matter fields. 

Since we are mainly interested in the effects of a gas of brane winding modes and transverse fluctuations on the evolution of a spatially homogeneous Universe, we will neglect the antisymmetric tensor field $B_{\mu \nu}$. We will use conformal time $\eta$
and take the background metric to be given by
\begin{equation}
G_{\mu \nu} \, = \, a(\eta)^2 diag(-1, 1, ..., 1) \, ,
\end{equation}
where $a(\eta)$ is the cosmological scale factor.

If the transverse fluctuations of the brane are small (in the sense that the first term on the right hand side of (\ref{indmet}) dominates) and the gauge fields on the brane are small, then the brane action can be expanded as follows:
\begin{eqnarray} \label{actexp}
S_{brane} \, &=& \, T_p \int d^{p+1} \zeta a(\eta)^{p + 1} e^{-\phi} \nonumber \\
& & e^{{1 \over 2} tr log(1 + \partial_m \phi_i \partial_n \phi_i + a(\eta)^{-2} 2 \pi \alpha' F_{mn})} \nonumber \\
&=& \, T_p \int d^{p + 1} \zeta a(\eta)^{p + 1} e^{- \phi} \\
& & (1 + {1 \over 2} (\partial_m \phi_i)^2 - \pi^2 {\alpha'}^2 a^{-4} F_{mn}F^{mn}) \, . \nonumber
\end{eqnarray}
The first term in the parentheses in the last line corresponds to the brane winding modes, the second term to the transverse fluctuations, and the third term to brane matter.  We see that, in the low energy limit, the transverse fluctuations of the 
brane are described by a free scalar field action, and the longitudinal fluctuations are given by a Yang-Mills theory. The induced equation of state has pressure $p \geq 0$.

The above result extends to the case of large brane field and brane position fluctuations. It can be shown \cite{Taylor} that large gauge field fluctuations on the brane give rise to the same equation of state as momentum modes ($E \sim 1/R$) and are thus
also described by pressure $p \geq 0$. In the high energy limit of closely packed branes, the system of transverse brane fluctuations is described by a strongly interacting scalar field theory \cite{Townsend} which also corresponds to pressure $p \geq 0$.

We will now consider a gas of branes and determine the equations of state corresponding to the various modes. The procedure involves taking averages of the contributions of all of the branes to the energy-momentum tensor, analogous to what is usually done
in homogeneous cosmology generated by a gas of particles.

Let us first focus on the winding modes. From (\ref{actexp}) it immediately follows that the winding modes of a p-brane give rise to the following equation of state:
\begin{equation} \label{EOSwind}
\tilde{p} \, = \, w_p \rho \,\,\, {\rm with} \,\,\, w_p = - {p \over d}
\end{equation}
where $d$ is the number of spatial dimensions (9 in our case), and where $\tilde{p}$ and $\rho$ stand for the pressure and energy density, respectively.

Since both the fluctuations of the branes and brane matter are given by free scalar fields and gauge fields living on the brane (which can be viewed as particles in the transverse directions extended in brane directions), the corresponding equation of state is that of ``ordinary" matter with
\begin{equation} \label{EOSnw}
\tilde{p} \, = \, w \rho \,\,\, {\rm with} \,\,\, 0 \leq w \leq 1 \, .
\end{equation} 
Thus, in the absence of a scalar field sector living on the brane, the energy will not increase as the spatial dimensions expand, in contrast to the energy in the winding modes which evolves according to (as can again be seen immediately from (\ref{actexp}))
\begin{equation} \label{winden}
E_p(a) \, \sim \, T_p a(\eta)^p \, ,
\end{equation}
where the proportionality constant depends on the number of branes. Note that the winding modes of a fundamental string have the same equation of state as that of the winding modes of a 1-brane, and the oscillatory and momentum modes of the string obey the equation of state (\ref{EOSnw}). 

In the context of a hot, dense initial state, the assumption that the brane
fluctuations are small will eventually break down. Higher order terms in the expansion of the brane action will become important. One interesting effect
of these terms is that they will lead to a decrease in the tension of the branes \cite{MR99}. This will occur when the typical energy scale of the system approaches the string scale. At that point, the state of the system will be dominated by a gas of branes.

The background equations of motion are \cite{TV,Ven}
\begin{eqnarray} \label{EOMback1}
- d \dot{\lambda}^2 + \dot{\varphi}^2 \, &=& \, e^{\varphi} E \\
\label{EOMback2}
\ddot{\lambda} - \dot{\varphi} \dot{\lambda} \, &=& \, {1 \over 2} e^{\varphi} P \\
\label{EOMback3}
\ddot{\varphi} - d \dot{\lambda}^2 \, &=& \, {1 \over 2} e^{\varphi} E \, ,
\end{eqnarray}
where $E$ and $P$ denote the total energy and pressure, respectively,
\begin{equation}
\lambda(t) \, = \, log (a(t)) \, ,
\end{equation}
and $\varphi$ is a shifted dilaton field which absorbs the space volume factor
\begin{equation}
\varphi \, = \, 2 \phi - d \lambda \, .
\end{equation}
In our context, the matter sources $E$ and $P$ obtain contributions from all components of the brane gas:
\begin{eqnarray}
E \, &=& \, \sum_p E_p^w + E^{nw} \nonumber \\
P \, &=& \, \sum_p w_p E_p^w + w E^{nw} \, ,
\end{eqnarray}
where the superscripts $w$ and $nw$ stand for the winding modes and the non-winding modes, respectively. The contributions of the non-winding modes of all branes have been combined into one term. The constants $w_p$ and $w$ are given by (\ref{EOSwind}) and (\ref{EOSnw}). Each $E_p^w$ is the sum of the energies of all of the brane windings with fixed p.

\vspace{0.4cm}
\section{Brane Gases in the Early Universe}

The first important conclusion of \cite{BV} was that in the approach to
string cosmology based on considering string gases in the early Universe, the initial cosmological (Big Bang) singularity can be avoided. The question we will now address is whether this conclusion remains true in the presence of branes with $p > 1$ in the spectrum of fundamental states.

The two crucial facts leading to the conclusions of \cite{BV} were t-duality and the fact that in the micro-canonical ensemble the winding modes lead to positive specific heat, leading to {\it limiting} Hagedorn temperature \footnote{The Hagedorn temperature is not reached at finite energy density \cite{ABKR}.}.
Both of these facts extend to systems with branes. First, as is obvious, each brane sector by itself preserves t-duality. Secondly, it was shown in \cite{ABKR} that if two or more Hagedorn systems thermally interact, and at least one of them (let us say System 1) has limiting
Hagedorn temperature, then at temperatures close to the Hagedorn temperature of System 1, most energy flows into that system, and the joint system therefore also has limiting Hagedorn behavior. Hence, as the Universe contracts, the t-duality fixed point $R = 1$ is reached at a temperature $T$ smaller than the Hagedorn
temperature, and as the background space contracts further, the temperature starts to decrease according to (\ref{tempdual}). There is no physical singularity as $R$ approaches 0.

Let us now turn to the dynamical de-compactification mechanism of 3 spatial dimensions suggested in \cite{BV}. We assume that the Universe starts out
hot, small and in thermal equilibrium, with all spatial dimensions equal (near the self-dual point $R = 1$). In this case, in addition to the momentum and oscillatory modes, winding modes of all p-branes will be excited. By symmetry, it is reasonable to assume that all of the net winding numbers cancel, i.e. that there are an equal number of winding and anti-winding modes.

Let us assume that the Universe starts expanding symmetrically in all directions. As $\lambda$ increases, the total energy in the winding modes increases according to (\ref{winden}), the contribution of the modes from the branes with the largest value of $p$ growing fastest. In
exact thermal equilibrium energy would flow from the winding modes into non-winding modes. However, this only can occur if the rate of interactions of the winding modes is larger than the Hubble expansion rate.

Generalizing the argument of \cite{BV}, from a classical brane point of view it follows (by considering the probability that the world-volumes of 2 p-branes in space-time intersect) that the winding modes of p-branes can interact in at most $2p + 1$ large \footnote{Large compared to the string scale.} spatial dimensions. Thus, in $d = 9$ spatial dimensions, there are no obstacles to the disappearance of $p=8$, $p=6$, $p=5$ and $p=4$ winding modes, whereas the lower dimension brane winding modes will allow a hierarchy of dimensions to become large. Since for volumes large compared to the string volume the energy of the branes with the largest value of $p$ is greatest, the 2-branes will have an important effect first. They will only allow 5 spatial dimensions to become large. Within this distinguished $T^5$, the 1-brane winding modes will only allow a $T^3$ subspace to become large.
Thus, it appears that the mechanism proposed in \cite{BV} will also apply if the Hilbert space of states includes fundamental branes with $p > 1$.

To what extent can these classical arguments be trusted? It has been pointed out in \cite{KKS} that the microscopic width of a string increases logarithmically as the energy with which one probes the string increases. However, in our cosmological context, 
we are restricted to energy densities lower than the typical string density, and thus the effective width of the strings is of string scale \cite{KKS}. Similar conclusions will presumably apply to branes of higher dimensionality. However, no definite results are known since a rigorous quantization scheme for higher dimensional branes is lacking. 

The cosmological scenario we have in mind now looks as follows: The Universe starts out near the self-dual point as a hot, dense gas of branes, strings and particles. The Universe begins to expand in all spatial directions as described by the background equations of motion (\ref{EOMback1} - \ref{EOMback3}). As space expands and cools (and the brane tension therefore increases), the branes will eventually fall out of thermal equilibrium. The branes with the largest value of p will do this first. Space can 
only expand further if the winding modes can annihilate.
This can be seen immediately from the background equations of motion (\ref{EOMback2}). If the equation of state is dominated by winding modes (which it would be if the Universe would keep on expanding), then (with the help of (\ref{EOSwind}) and (\ref{winden})) it follows that the right hand side of that equation acts as if it comes from a confining potential
\begin{equation} \label{pot}
V_{eff}(\lambda) \, = \, \beta_p e^{\varphi} e^{p \lambda} \,
\end{equation}
where $\beta_p$ is a positive constant which depends on the brane tension.

The unwinding of p-branes poses no problem for $p=8$, $p=6$, $p=5$ and $p=4$ \footnote{For a discussion of the microphysics of brane winding mode annihilation see e.g. \cite{Sen} and references therein.}. The corresponding brane winding modes will disappear first. However, the $p = 2$ branes will then only allow 5 spatial dimensions to expand further (which 5 is determined by thermal fluctuations). In this distinguished $T^5$, the one-branes and fundamental strings will then only allow a $T^3$ subspace to expand.
Thus, there will be a hierarchy of sizes of compact dimensions. In particular, there will be 2 extra spatial dimensions which are larger than the remaining ones. The connection with the large-extra dimension scenario proposed in \cite{ADD} is intriguing. 
However, since the only fundamental scale in the problem is the string scale, and since there does not seem to be a way to naturally separate the temperature scales at which the various branes fall out of equilibrium, it appears difficult to produce the mm scale proposed in \cite{ADD}.

Note that even when winding mode annihilation is possible by dimension counting, causality imposes an obstruction. There will be at least 1 winding mode per Hubble volume remaining (see e.g. \cite{Kibble,RBrev}). In our four-dimensional space-time, all branes with $p \geq 2$ will look like domain walls. This leads to the well known domain wall problem \cite{problem} for cosmology since one wall per Hubble volume today will overclose the Universe if the tension of the brane is larger than the electroweak scale. Due to their large tension at low temperatures, even one-branes will overclose the Universe.

There are two ways to overcome this domain wall problem. The first is to invoke cosmic inflation \cite{Guth} at some stage after the branes have fallen out of equilibrium but before they come to dominate the energy of the Universe. The scenario would be as follows: initially, the winding modes dominate the energy density and determine the dynamics of space-time. Once they fall out of equilibrium, most will annihilate and the energy in the brane winding modes will become subdominant. The spatial dimensions
in which unwinding occurs will expand. It is at this stage that the ordinary field degrees of freedom in the theory must lead to inflation, before the remnant winding modes (one per Hubble volume) become dominant.
 
In our context, however, there is another and possibly more appealing alternative - loitering \cite{SFS}. If at some stage in the Universe the Hubble radius \footnote{More precisely, the ``causal horizon'', meaning the distance which light can travel during the time when $H^{-1}$ is larger than the spatial extent.}  becomes larger than the spatial extent of the Universe, there is no causal obstruction for all winding modes to annihilate and disappear. In the context of ``standard" cosmology (fields as matter sources coupled to classical general relativity) it is very hard to obtain loitering. However, in the context of dilaton cosmology a brief phase of loitering appears rather naturally as a consequence of the confining potential (\ref{pot}) due to the winding modes. To see this, let us return to the background equations (\ref{EOMback1} - \ref{EOMback3}). The phase space of solutions was discussed for general values of $p$ in \cite{TV}. 

For a fixed value of $p$, the phase space of solutions is two-dimensional and is spanned by $l = {\dot \lambda}$ and $f = {\dot \varphi}$. If we start in the energetically allowed part 
\begin{equation}
\vert l \vert \, < \, {1 \over {\sqrt{d}}} \vert f \vert 
\end{equation}
of the upper left quadrant of phase space with $l > 0$ and $f < 0$ corresponding to expanding solutions with ${\dot \phi} < 0$, then the solutions are driven towards $l = 0$ at a finite value of $f$ (see Figure 1). More precisely, there are three special lines in phase space with \cite{TV}
\begin{equation}
{{\dot l} \over {\dot f}} \, = \, {l \over f} \, ,
\end{equation}
which correspond to straight line trajectories through the origin. They are
\begin{equation}
{l \over f} \, = \, \pm {1 \over {\sqrt{d}}} \, , \, {p \over d} \, .
\end{equation}
Solutions in the energetically allowed part of the upper left quadrant are repelled by the special line 
$l/f = - 1/\sqrt{d}$ and approach $l = 0$. For $l \rightarrow 0$, both $\dot{l}$ and $\dot{f}$ remain finite
\begin{equation}
{\dot l} \, \simeq \, - {p \over {2d}} f^2 \, , \, {\dot f} \, \simeq \, {{f^2} \over 2} \, .
\end{equation}
Hence, the trajectories cross the $l = 0$ axis at some time $t_1$. This means that the expansion of space stops and the Universe begins to contract. The crossing time $t_1$is the first candidate for a loitering point.

Note that the dynamics of the initial collapsing phase is very different from the time reverse of the initial expanding phase. In fact, as the special line $l/f = p/d$ is approached, ${\dot l}$ changes sign again, and the trajectory approaches the static 
solution $(l, f) = (0, 0)$ - also implying a fixed value for the dilaton. At late times, the time evolution along the abovementioned special line corresponds to contraction with a Hubble rate whose absolute value is decreasing,
\begin{equation}
H(t) \, = - {1 \over {\vert H^{-1}(t_0) \vert + \beta (t - t_0)}} \, ,
\end{equation}
(where $t_0$ is some starting time along the special line) with 
\begin{equation}
\beta =  {p \over 2} + {d \over {2p}} \, .
\end{equation}
Thus, the evolution slows down and loitering is reached, even if the time
evolution at $t_1$ is too rapid for loitering to occur then. Note that these considerations assume that the winding modes are not decaying. If they decay too quickly, this could obviously prevent loitering.
 
\begin{figure}[htbp]
\includegraphics[angle=270,width=3.5in]{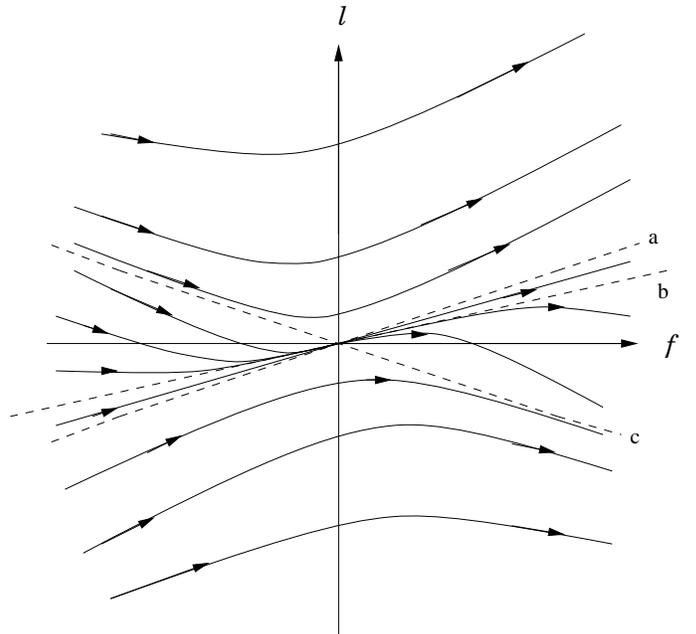}
\caption {Phase space trajectories of the solutions of the background equations (\ref{EOMback1} - \ref{EOMback3}) for the values $p = 2$ and $d = 9$. The energetically allowed region lies near the $l = 0$ axis between the special lines {\it a} and {\it c}, which are the lines given by $l/f = \pm 1 / \sqrt{d}$. The trajectory followed in the scenario investigated in this paper starts out in the upper left quadrant close to the special line {\it c} (corresponding to an expanding background), crosses the $l = 0$ axis at some finite value of $f$ (at this point entering a contracting phase), and then approaches the loitering point $(l, f) = (0, 0)$ along the phase space line {\it b} which corresponds to $l/f = p/d$.}
\label{fig1}
\end{figure}

Note that given our cosmological starting point there is no horizon problem since space was initially of string size. However, two of the other problems of standard cosmology which the inflationary scenario successfully addresses, namely the flatness and the formation of structure problem, persist in our scenario. Note, in particular, that it seems necessary to have something like inflation of the large spatial dimensions in order to produce a Universe which is larger than the known Hubble radius. It is of great interest to explore the possibility of finding a solution of these problems in the context of string theory, e.g. along the lines of the string-driven inflationary models of \cite{Turok,Abel}.
 
\vspace{0.4cm}
\section{Discussion and Conclusions}

In this paper we have generalized the approach to superstring cosmology pioneered in \cite{BV} to include the contribution of branes. Our starting point is the assumption that the Universe starts out small, hot and in thermal equilibrium, with all of the spatial dimensions being equivalent and compact (string scale). We also assume that the background admits one cycles. We work in the corner of moduli space resulting from compactification of 11-d M-theory on $S^1$, and in which the string coupling constant is small. 

We argue that, as a consequence of t-duality, the usual big bang singularity is absent in the resulting cosmology. Furthermore, since winding modes of all of the branes are excited in thermal abundance in the initial state, and since the energy in winding modes increases as the background spatial scale expands, the thermodynamics of the winding modes coupled to the background equations of motion, which in the corner of moduli space of M-theory which we consider are the dilaton-gravity equations, dominates the initial evolution of the background. Our thermodynamic considerations suggest that the mechanism first pointed out in \cite{BV}, by which the string winding modes will prevent all but 3 spatial dimensions from becoming large, persists. In addition, the presence of winding modes of 2-branes may lead to a hierarchy in the sizes of the extra dimensions, with exactly internal dimensions being larger than the others.

We pointed out some further interesting characteristics of the resulting cosmology. In particular, the horizon problem is absent. However, in order to produce a Universe in which the large spatial dimensions exceed the present Hubble radius, it seems necessary to have a background evolution of the large spatial dimensions resembling inflation. 

We also pointed out the existence of a {\it brane problem}, a problem for cosmology in theories with stable branes which is analogous to the domain wall problem in cosmological scenarios based on quantum field theories which admit stable domain walls. A phase of loitering in the background cosmological evolution will naturally solve this problem. Based on the background equations of motion, it appears that as long as the winding modes do not disappear, the background solutions approach a point of loitering.

It would be interesting to extend our considerations to other regions in moduli space, in particular to regions of strong string coupling. In those regions it appears \cite{MR99}that the effective ten-dimensional background description breaks down before the Hagedorn temperature is reached.

It is also important to point out that on Calabi-Yau threefold backgrounds one cycles are absent, and thus our cosmological scenario does not apply. Calabi-Yau three-folds are required if the four-dimensional low energy effective theory is to have $N=1$ supersymmetry. In the context of early Universe cosmology, however, it is not reasonable to demand $N=1$ supersymmetry. In particular, one could have maximal supersymmetry which is consistent with the toroidal background we are using. It would be interesting to explore whether consistent four-dimensional low-energy effective theories can be constructed from compact backgrounds which admit one-cycles. 

Given that the tension of the branes exceeds that set by the string scale, the applicability of the homogeneous background field equations to brane gases might be questionable, in particular at late times when the branes are fairly widely separated \footnote{We thank D. Lowe for stressing this concern to us.}. This issue deserves further study.

\vspace{0.4cm}
\centerline{\bf Acknowledgements}

The research was supported in part by the U.S. 
Department of Energy under Contract DE-FG02-91ER40688, TASK A. Two of us
(S.A. and D.E.) are supported by fellowships by the U.S. Department of Education under the GAANN program. We are grateful to E. Akhmedov, R. Easther, B. Greene, A. Jevicki, D. Lowe and S. Ramgoolam for many discussions about this project.

\end{document}